# Review on Computer Vision in Gastric Cancer: Potential Efficient Tools for Diagnosis

Yihua Sun[1]

[1]School of Mathematics (Zhuhai), Sun Yat-sen University, China

Corresponding author: Yihua Sun (e-mail: sunyh9@mail2.sysu.edu.cn).

**ABSTRACT** Rapid diagnosis of gastric cancer is a great challenge for clinical doctors. Dramatic progress of computer vision on gastric cancer has been made recently and this review focuses on advances during the past five years. Different methods for data generation and augmentation are presented, and various approaches to extract discriminative features compared and evaluated. Classification and segmentation techniques are carefully discussed for assisting more precise diagnosis and timely treatment. For classification, various methods have been developed to better proceed specific images, such as images with rotation and estimated real-timely (endoscopy), high resolution images (histopathology), low diagnostic accuracy images (X-ray), poor contrast images of the soft-tissue with cavity (CT) or those images with insufficient annotation. For detection and segmentation, traditional methods and machine learning methods are compared. Application of those methods will greatly reduce the labor and time consumption for the diagnosis of gastric cancers.

## I. INTRODUCTION

Gastric cancer is one of the most severe tumors with high mortality. Rapid and accurate diagnosis of gastric cancer or the risk of gastric cancer through multiple imaging processes contributes greatly to prompt treatments. However, it is not only time consuming to analyze the patients' gastric images, but also a risk of missing important images because sometimes lesion images differ only moderately from the normal ones. Fortunately, with the development of computer vision techniques and the improvement of computing power, it is possible to design proper methods to assist experts in analyzing medical images, enhancing the diagnosis accuracy and efficiency. Besides, with help of computer vision techniques, less manpower in the field of gastric cancer and other medical fields will be needed. Recently, computer vision has been rapidly developed and shown high performance in natural picture processing. Currently, lesion classification, and cancer region detection and segmentation are the mostly studied methods.

Lesion classification identifies the kind of gastric lesions, i.e. classifying the pictures into different categories: lesion, cancer or normal. Lesion classification is an image-level categorizing task that helps doctors to rapidly screen patients' gastric images. Cancer region detection and segmentation is the extension analysis of lesion classification that locates the regions of lesions or cancers in a given image. To conduct region detection, a bounding box is set around the lesion or cancer in the image. As to the segmentation, the exact boundary of the lesion or cancer is identified, and in other words, segmentation can be viewed as a pixel-level classification task that identifies whether each pixel in the image belongs to cancer region or normal region. The lesion region detection and segmentation will provide a more detailed information for doctors, helping them to better analyze medical images and even to remove cancer tissues more precisely during operation. There are also some less studied tasks that may help doctors' diagnosis to some extent, including counting lymph nodules, tracking cancer cells, providing panoramic view of stomach, cell nuclei segmentation and 3D reconstruction, etc.

Importantly, feature extraction and classification are two critical components and steps for a successful process of a specific clinical image data. Here, papers of past five years (2015-2019) on computer vision in gastric cancer focusing on developing medical image processing techniques are reviewed. The papers were enrolled through searching "(gastric cancer) AND (computer vision)" as keywords on Web of Science, Scopus, and PubMed. Then the papers that use computer vision method to analyze gastric image data were chosen. To compare the methods, matrices including sensitivity (SEN), specificity (SPE), precision (PRE), accuracy (ACC), F1-score (FSC), receiver operating characteristic (ROC) curve, area under curve (AUC), etc., are calculated and all the works are summarized in tables. If a paper compares the performance of different approaches, only the best approach is listed in the tables. The portion of the papers in each year reviewed in this paper is about the same as the portion of publications per year in this field. There is a growing intendency in applying computer vison techniques to assist diagnosing gastric cancer recently (Fig. 1).

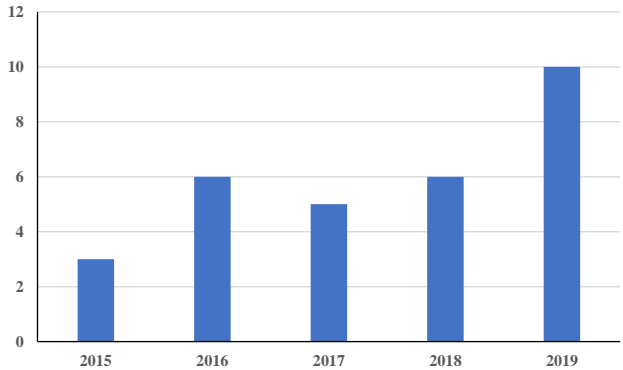

**FIGURE 1.** number of publications in each year reviewed in this paper.

## II. LESION CLASSIFICATION

Due to the insufficiency of images with lesions, generative adversarial networks (GAN) were implemented to generate fake images that share the same characters with real ones to boost the strength of classification [1]. Besides, the classification problem can be formulated as a multi-instance learning (MIL) problem to avoid additional annotation [2, 3]. The reason lies in that when a patient is diagnosed with a disease based on image appearance, there must be at least one abnormal image in the series of images from the patient. Various researches based on convolutional neural network (CNN) have been done [2, 4-11]. Gabor filter and gray level co-occurrence matrix (GLCM) are commonly utilized to capture texture information [12-16]. In addition, the classifier support vector machine (SVM) always reaches the best performance among other classifiers [6, 12, 14-17]. Endoscopy examinations and pathological image analysis, which have attracted more research interests, are most important for making diagnosis in clinical practice, (Fig. 2).

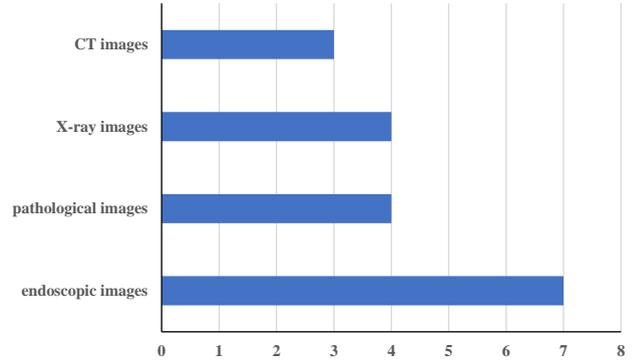

**FIGURE 2.** number of publications of each image type reviewed in this section.

### A. ENDOSCOPY

Endoscopy is one of the most common examination tools for early gastric cancer and gastritis in clinical practice. There are two main kinds of endoscopy: standard imaging endoscopy and image-enhanced endoscopy. The difference lies in the application of several techniques of image-enhanced endoscopy with the purpose to enhance texture information captured by endoscopy, such as, chromoendoscopy by spraying a specific dye and magnification endoscopy by using narrow-band spectrum filter.

For the fact that more discriminative texture information can be obtained by image-enhanced endoscopy, more research on which is based on texture analysis with hand-crafted features. One of the characteristics of endoscopy images is that the lesion area is always in a circle or an oval. Besides, as the endoscope may rotate during the examination, features should be rotation-tolerant [18]. In terms of the limitation of medical dataset, transfer learning also draws attention to develop more powerful models with endoscopy [5, 11]. The papers are summarized in Table I.

#### 1) STANDARD IMAGING ENDOSCOPY

As endoscopy images are always captured from different angles and the lesions have a round shape, Zhao, et al. [18] designed a rotation-tolerant image feature extraction method, TriZ, to classify lesion images. The images were sliced into several rectangular ring layers by rectangular boxes of different sizes with the same aspect ratio as the original images. The images were summarized as histograms with respect to the gradient and orientations of each pixel. TriZ [18] with relatively less extracted features from histograms outperformed other feature extraction methods. Besides, random forest (RF) with TriZ outperformed other classifiers.

Sakai, et al. [11] produced a heatmap probability for early gastric cancer. They randomly cropped 100 patches (224× 224), in which over 80% of cancerous regions were included, from each of the selected cancer images to form the cancer image training set. The normal image training set was formed by cropping patches randomly from normal region in cancer images as well as completely from normal images. In the following step, the parameters of GoogLeNet [19] pretrained on ImageNet were fine-tuned to obtain a transfer learning classifier. In terms of detecting lesions, they cropped partly overlapped patches from the endoscopy images and obtained a heatmap to indicate the probability of having early gastric cancer in the patches, so as to assist doctors with useful information for diagnosis.

Based on the idea of asymmetric principle component analysis (APCA) and singular value decomposition (SVD), Liu, et al. [17] proposed joint diagonalization principle component analysis (JDPCA) that could reduce the dimension and solved the overfitting problem to avoid loss of discriminative information on an unbalanced dataset. JDPCA calculated the class-conditional covariance matrix of both positive and negative samples. The features of the images were extracted by JDPCDA in the YCbCr color space and the color coherence vector (CCV) in the RGB space. An improved SVM, ODR-BSMOTE-SVM (OB-SVM) [20], was implemented on the features to finally classify the images. Besides gastric images, the authors also tested the method on esophagus images and received expected results, indicating that JDPCA could be applied to various data other than on a specific type of data.

2) IMAGE-ENHANCED ENDOSCOPY

Ali, et al. [12] classified normal and tumor images with hand-crafted texture features and well-known classifiers. They implemented 2D Gabor filter with 8 different scales and 8 different orientations to perform a multi-resolution spatial-frequency analysis on the images. After every input image was filtered with the Gabor filters, the multi-scale responses were followed by the GLCM to generate Gabor-based gray level co-occurrence matrix (G2LCM) to analyze the images. The authors finally trained 6 different classifiers on the features extracted by 3 different methods. The results showed that the performance of the SVM classifier trained on the texture extracted with G2LCM was the best.

Boschetto and Grisan [13] firstly segmented superpixels, which were clusters of connected pixels with similar texture, by simple linear iterative clustering (SLIC). Statistics including mean intensity and standard deviation, GLCM, and local binary patterns (LBP) were employed to extract 111 features from each superpixel. Based on the features and random forest classifier, the superpixels were classified to be normal or abnormal. The final classification of the whole image was carried out by the majority vote algorithm of the superpixels. The accuracy in the image-level was higher than the accuracy in the superpixel-level, and the result in the superpixel-level could be regarded as a kind of gastric cancer region detection.

Rather than the usually mean and deviation that may not differ significantly in different cases, Ali, et al. [15] calculated the geometric mean and geometric standard deviation (geometric homogeneous texture, GHT) of the responses to extract features from the original images. Then the metaheuristic-based genetic algorithm (GA) with higher accuracy was applied to implement feature selections that discarded redundant features. The final results revealed that GA could not only reduce the dimensionality, but also increased the accuracy. The result also indicates that GHT features are more discriminative, thus outperforming the traditional LBP. Meanwhile, SVM performed as the best classifier on the proposed features.

Liu, et al. [5] implemented the transfer learning to compensate for the unavailability of large dataset to train a deep learning network. Serval CNNs pretrained on the large natural image dataset were taken into consideration. What is more, Inception-V3 [21] pretrained on ImageNet was considered to be the best CNN for the following experiments. They compared different transfer learning scenarios including fine-tuning all layers, front half of the layers, and only fully-connected layers. It was shown that all layers fine-tuned network had the best performance, which can be ascribed to the difference of the low-level features of magnifying endoscopy with narrow-band imaging (M-NBI) images from the natural ones. Moreover, classified by SVM, the transfer learning method outperformed traditional machine learning methods with handcraft features extracted by LBP, complete local binary pattern (CLBP), Gabor filter and GLCM.

B. PATHOLOGICAL IMAGES

Pathological images are important for gastric cancer diagnosis, whereas the diagnosing process is error prone and time consuming for pathologists. Those images are always in a high resolution, increasing the complexity of models. To train the model more efficiently, small scale patches are usually cropped from the original high-resolution images. Additionally, detailed annotation of cancerous regions is laborious. Under most circumstances, only the image-level annotation can be obtained. In this case, weakly supervised approaches are required to develop a patch-based model without the patch-level annotation. Besides, to reduce the model complexity and the execution time, a teacher-student transfer learning framework is usually implemented [8]. The papers are summarized in Table II.

Li, et al. [4] proposed a CNN based on a deep learning framework. This method used patch-based results to proceed the slice-based final classification. They first cropped patches with a 224×224 window from the original image by sliding the window over the whole image of normal slices and locating the window over the cancer area of cancerous slices. Their method consisted of two parts: the patch-based classification and the slice-based classification. In terms of the patch-based classification part, they proposed a CNN based framework with multi-scale module (MSM) in shallow layers and network in network module (NIN) in deep layers. The MSM helped to detect multi-scale targets in gastric slices, while the NIN fused the features extracted by MSMs with fewer parameters. In this part, the model output the probability of being a normal patch. As to the slice-based classification part, they averaged the lowest 10 scores of all the patches in a complete slice as the possibility of cancer. With a pre-set threshold as 0.2, they finally determined whether the slice contained cancer. The patch-based classification reached a high accuracy of 0.9793 and the final slice-based classification had an accuracy of 100%. The authors showed that their method outperformed the well-known AlexNet [22], DenseNet [23], ResNet [24], etc. However, as they did not present the figure of ROC, their pre-set threshold might only fit this specific BOT dataset. More experiments are required to test this method on other datasets.

As most clinical data are weakly annotated, usually only in image-level, it is necessary to develop methods to analyze those data, especially those obtained in early years. To solve this problem, Li, et al. [10] designed multilayer hidden conditional random fields (MHCRF) framework that could make use of both traditional feature extraction methods and machine learning feature extraction methods to classify cancerous histopathological images with only image-level labels. They first extracted features with hand-crafted RGB gray histogram, scale-invariant feature transform (SIFT) [25], DAISY, GLCM [26], histogram of oriented gradient (HOG), state-of-art Inception-V3 [21], as well as VGG-16 [27]. Then the features were classified with SVMs, artificial neural network (ANN), and random forest (RF). Then the best combinations of the patch-level feature-classifier were selected and used jointly for the image-level classification. Also, the joint probability was selected to generate the unary and binary potentials of the images. Finally, the MHCRF that could utilize sequential and context information in images was trained based on the above potentials to carry out the final image-level classification, thus further increasing the accuracy. The authors implemented and analyzed the proposed method not only on gastric histopathological images but also on cervical histopathological images. Although this method performed relatively better on the gastric dataset than on cervical images, it still showed the potential for extrapolation to other fields.

Sharma, et al. [7] developed a self-designed CNN for cancer and necrosis classification. The images were annotated by 5 experts and the intersections marked by most experts were selected and then augmented. Regarding the cancer classification, the accuracy of non-tumor was the as high as 0.8203, while the accuracy of HER2- tumor was the lowest with 0.5809, which could attribute to the less or no immunohistochemical responses from cells in the tumor area. As to the necrosis classification, the proposed CNN outperformed AlexNet [22] and other handcrafted methods. The higher ability of distinguishing necrosis than that of tumor led to the higher accuracy of necrosis classification based on even fewer data. The sharp contrast of necrotic tissues from normal or tumorigenic tissues can be a possible reason. However, this CNN was inefficient in training and an extension of data with more patients required to improve the robustness.

When it comes to practical application of cancer classification, besides accuracy, the execution time of a model to classify an image should be taken into account, especially when processing the high-resolution whole slide images (WSI). Kong, et al. [8] tried to solve this problem by model compression that trained a small efficient "student" network with fewer parameters while having the ability of accessing large "teacher" network with high accuracy. A small network utilizing depthwise separable convolution was regarded as the "student" network, which reduced a large amount of computation, compared to the conventional convolution. The large network Inception-V3 [21] was used as the "teacher" network that had been pretrained on the patches of the original WSI and reached high accuracy. The "student" network was then trained on the above patches again with a customized loss function that guided the "student" network to learn the features (intermediate layers) and output of the "teacher" network. To further reduce the memory cost of the small network, the authors transformed the fully connected layers to full convolution networks (FCN). In addition, FCN allowed the network to handle arbitrary size of images including WSI. As a result, a probability map of the input WSI was obtained and used for the classification. In terms of the average Free-response ROC, the performance of the proposed small network was almost the same as that of the large inception-V3, while the processing time of that small network was about 5 times faster.

### C. X-RAY

Although the diagnostic accuracy of the X-ray examination is relatively lower than that of the endoscopy of gastric cancer, X-ray is more efficiency for quick diagnosis compared with endoscopy. In addition, X-rays are significant for gastritis diagnosis for they can alarm the risk to have gastric cancer. Perhaps as the accuracy of the X-ray diagnosis increases with the help of computer vision techniques, X-ray may contribute more in fast and accurate diagnosis in clinical practice. Same as other medical images with insufficient datasets, GAN is a promising idea to address this problem. The papers are summarized in Table III.

Togo, et al. [1] proposed the loss function-based conditional progressive growing GAN (LC-PGGAN) to solve the problem of insufficient image data. They first cropped patches from the original X-ray images. Then, fake images were generated by progressive growing GAN (PGGAN), which started with low resolution images and progressively increased the resolution of image generation by adding new layers to the generator and the discriminator. As a result, PGGAN could generate high resolution images as well as avoiding model collapse. Moreover, the authors adjusted the loss function in PGGAN by enabling the model to process conditional information, so that the LC-PGGAN could judge the generated gastritis, non-gastritis, outside images as well as fake images differently. In order to evaluate the quality of the generated images, the authors implemented GAN-train to test the classifier trained on the fake images with real image testing dataset. Although the performance of SVM trained on the images generated by LC-PGGAN outperformed those generated by PGGAN and DCGAN, it could not catch up with that trained on real images.

Ishihara, et al. [6] developed a gastric cancer risk detection method from a single X-ray image labeled with clinical examination. The authors separated the entire images into patches and developed a patch-based CNN model. After the relevance score was calculated according to the CNN output, a number of patches most related to the image-level label were selected. In the following, the CNN was retrained with the selected patches for the work afterwards. To adjust to the different size and shape of patients' gastric images, the authors implemented bag of features representation (BoF) on the output value of the CNN intermediate layer to extract representative features. Eventually, whether the whole images suggest the risk of cancer was determined by the features via SVM. The comparison showed that this method outperformed the handcraft feature-based method and the doctors' performance.

Togo, et al. [16] improved the accuracy for classifying gastritis images by estimating the salient related to the chronic gastritis. One contribution map of patches was calculated by the SVM classifier, suggesting the relation to gastritis and non-gastritis. Another contribution map used to remove the regions outside the stomach. As a result, the salient regions related to the chronic gastritis or non-gastritis were obtained with a threshold to the dot product of the above maps. The results showed the accuracy of image-level classification was improved only with the information in the estimated salient regions.

Ishihara, et al. [14] proposed a method to classify X-Ray images of helicobacter pylori infection that counted for the gastric cancer risk. They extracted several hand-crafted features by employing intensity histogram, Gabor Wavelet and other methods. On this basis, they used a feature selection method, minimum-redundancy maximum-relevance (mRMR) algorithm, to reduce the dimension of features. The authors utilized SVM and multiple kernel learning (MKL) that could optimize the combination of kernels and parameters to classify the images. The multiple results carried out by the classifiers were integrated by late fusion where expectation maximization (EM) algorithm outperformed major voting.

### D. COMPUTED TOMOGRAPHY (CT)

CT is useful for preoperative assessment. However, to make a diagnosis of gastric cancer is full of challenge because of the complexity around the stomach, the poor contrast of the soft-tissue with cavity, and the high cost of annotating each CT image. It can be concluded that at least one image contains cancerous region in the whole CT volume of a patient with cancer. Hence, viewing the classification challenge as a MIL problem, a weakly supervised method, is proper for CT images. Beyond that, a good model can also be trained with the help of expert experience. The papers are summarized in Table IV.

By identifying the cancerous multiple layer CT images, one could obtain the tumor invasion depth information. Li, et al. [3] formulated this classification task as a MIL task, in which each image was a bag and patches extracted from the image were instances. They first extracted the patches with initial seed points in both suspected and normal regions in the gastric wall to ensure the existence of positive instances in the positive bags. Next, they extracted bag-level features as well as instance-level features for better representation. To obtain the bag-level features, it was necessary to extract both statistical features including location, cross-sectional dimension, morphological characteristics, and spectral CT features including monochromatic CT values, material density measurements, effective-Z values. As to the instance-level features, the response of circular Gabor filter (CGF) was convoluted with a mask to extract texture features.

Fang, et al. [2] did similar study on the cancer classification as a generalized MIL task, they took a group of informative patches instead of a dissipative instance in a positive bag to adjust to the intra-class variation and inter-class ambiguity in the images of gastric cancer. Hence, they developed a CNN model trained by 2 stages, which could provide an accurate classification and a guidance to recognize other tissues in the image. The author showed that the final performance of the model outperformed both standard CNN model and a texture-based model with 90 extracted hand-craft texture features. Taking the problem as a MIL task is promising for analyzing medical images lacking of annotations.

Fang, et al. [9] also used a generic multi-task learning framework (MTL) to better learn some manually selected representations. The image patches in the tumor center were marked and extracted. A feature vector of each patch was obtained by a CNN and input to 2 distinctively and fully connected layers. For the first auxiliary task, the vector is used to predict selected auxiliary characteristics, including carcinoembryonic antigen level and clinical T stage. For the second major task, the vector is concatenated with the predicted characteristics to discriminate serosa invasion (T4 stage) in advanced gastric cancer. The model was trained through 2 strategies, starting with randomly initialized parameters or starting with model pretrained on ImageNet. The results showed that the MTL outperformed the commonly used single-task framework as well as the transfer learning strategy. The authors also implemented the MTL on lung cancer images to demonstrate its generalization.

## III. CANCER REGION DETECTION AND SEGMENTATION

Cancer region detection and segmentation provide more informative guidance for medical approach than the classification. More researches in this area are demanding as well. Besides insufficient images, annotation (ground truth) cost is higher in this task than in the classification. Consequently, GAN is also supposed to play its role in generating data along with proper annotation [28]. Weekly supervised method also helps reduce the annotation cost [29]. Machine learning methods [29, 30] have not yet shown their superiority over traditional methods [31-33]. The papers are summarized in Table V.

### A. CANCER REGION DETECTION

Since the number of patients with gastric cancer is small and the cost of the expert annotation is high, the problem of lacking training data always interferes the performance of the cancer detection model even with the data augmentation methods. The main contribution of Kanayama, et al. [28] is that they proposed a method based on GAN to enlarge the dataset of the annotated gastric lesion image with a small lesion image dataset, the size of which was 1% of the normal image. The researchers augmented their database by flipping, rotation, grayscale, and channel shuffle techniques. They generated the new cancer image by smoothly synthesizing normal gastric images with the lesion part from lesion images, so that the new image would have already been annotated with bounding box once it was generated. A discriminator, consisting of a global part focusing on the whole image and a local part focusing on a lesion area, was constructed to judge the generated images. The discriminator ensured that the generated images looked real and the lesion part synthesized smoothly. The proposed GAN generated lesion image more clearly than the well-known DCGAN. The authors then trained a cancer detection network with the enlarged dataset. According to the result, the enlarged dataset increased the performance of the network.

Ural, et al. [34] developed a gastric cancer regional detection system by analyzing the power spectrum density of processed CT images. The authors applied fast Fourier transform (FFT) to separate the real and imaginary part of an image. Then, the log transform was taken on the amplitude of FFT result to compress the lightness while expand the darkness. Finally, the power spectrum density (PSD) curve of the images was obtained and the cancerous regions corelated to recognizable frequency range in PSD curve. The locations of the normal and cancerous regions of 25 in 30 patients were detected successfully with the proposed method. The combination of FFT with the log transform also outperformed the methods separately employed these 2 algorithms.

### B. CANCER REGION SEGMENTATION

**Encoder-decoder system** Sun, et al. [30] have done lots of experiments on different network structures before proposing their segmentation method with high accuracy. With ResNetV2-101 pretrained on ImageNet, they constructed an encoder that fused different semantic-level features. Then they built a lightweight decoder with depthwise separable convolution and a residual connection. Finally, they implemented dense upsampling to refine the segmentation details. The authors claimed that their methods would perform better with a more powerful hardware to take full advantage of the batch normalization.

**Fully convolutional network (FCN)** The pixel-level bionomical image annotation relies heavily on specialists and costs lots of time. Therefore, there is a lack of training data for fully supervised learning. Liang, et al. [29] proposed a weakly supervised segmentation method based on FCN, which only required partially annotated images for training and cost less for data labeling. Their original dataset was provided by the 2017 China Big Data and Artificial Intelligence Innovation and Entrepreneurship Competition. They applied an overlapped region forecast method to avoid coarse predictions by ensuring that a complete boundary, e.g. a protrusion, was more likely to be included in a single patch. The key step was to iteratively train the model on the self-annotated dataset. They combined the high probability annotations presented by the model with the original labels. The model was then fine-tuned on the augmented dataset with 2 sampling methods. Nevertheless, their method could not completely avoid overfitting.

**Region Growing (RG)** Yasar, et al. [31] conducted some research on endoscopy image with region growing (RG) and statistical region merging (SRM) to segment cancerous regions. RG starts from a selected initial pixel and clusters the similar pixels in the neighbor to a same region iteratively. SRM is a colored image segmentation technique that starts with one region per pixel and applies statistical test on the neighboring region. In this case, similar regions will be merged iteratively. The authors also implemented RG on the area segmented by SRM, namely statistical region merging with region growing (SRMWRG), to recover the missing area of SRM. The ground truth was annotated by doctors and the pixel matching used to measure the performance of the methods. The final results showed that RG was the method with the best performance.

**Mueller matrix** Wang, et al. [32] investigated the potential of Mueller matrix for 3 category image segmentation (digital staining), cancer, dysplasia, and intestinal metaplasia/normal glands. Each region of interest (ROI) represented a point on the final digital stained images. Moreover, all the pixel information on the input data within a ROI was averaged as one point. Different sizes of ROI were investigated, in which case larger ROI yielded the lower resolution digital stained images. ROI of $1 \times 1$ could be viewed as the pixel level segmentation. Meanwhile, ROI of $640 \times 512$ (the size of the image) could be viewed as the image level classification. The authors finally decided to use ROI of 8×8 as a trade of between the accuracy and resolution. In each sample, 2 thin vertical sections next to each other were obtained, with one stained with hematoxylin and eosin (H&E) while the other unstained. The stained section helped locate the demanding imaging region of the corresponding unstained section. The unstained one was processed with the polarimetry imaging and the Mueller matrix obtained. After that, the unstained section was stained and analyzed by specialist for validation. They performed two-step classification that firstly distinguished the low-risk group (intestinal metaplasia/normal glands) from the high-risk group (cancer and dysplasia) and then classified cancer and dysplasia in the high-risk group. The author discovered that the selected Mueller matrix elements outperformed their derivatives like polarization parameter and principal component scores. Linear discriminant analysis was implemented to find the optimal combination of the elements. By classifying the ROIs, the image segmentation (digital staining) was obtained. Different size of ROI, 640×512, 8×8, and 1×1, were respectively corelated with different accuracies, 0.75, 0.59, and 0.56. The accuracy of identifying dysplasia is low, which may reflect the minor difference between the high and low grade of dysplasia.

**Chan-Vese (C-V) model and hue texture** Liu, et al. [33] proposed a hue-texture-embedded region-based model utilized the global and local information for the segmentation of gastric abnormalities, such as chronic gastritis, metaplasia, atrophy, neoplasia and the early gastric cancer. For the fact that the lesions differ in color and shape of microvessels, the authors extracted features from the color and the shape of microvessels for image segmentation. Hue transformation followed by Gaussian smooth filter was utilized to define the global hue energy functional representing color variations. Adaptive threshold on M-NBI saturation images was utilized to define the local microvascular texture energy functional representing local pixel similarity. They proposed a novel hue-texture-embedded joint energy with the combination of the above energy functional and a length regularization. With the framework of C-V model and hue-texture-embedded joint energy, the hue-texture-embedded region-based model was constructed and outperformed the C-V model. The method could handle different abnormalities, whereas it could be adequate only for specific M-NBI scenario.

## IV. OTHER LESS RESEARCHED BUT IMPORTANT QUESTIONS

The cell-level analyzation is time consuming when the cells distribute densely in an image, but helpful for diagnosis, such as analyzing lymph nodules [35] and cell nucleus [36]. Panoramic view [37] and 3D reconstructed model [38] provide broader view for doctors to make more comprehensive diagnosis. Cell tracking [39] helps to analyze cell movement. The papers are summarized in Table VI.

**Lymph nodule counting** Zhengdong, et al. [35] proposed a method based on faster R-CNN [40] transferred from ImageNet to efficiently detect and localize metastatic lymph in a single CT image, taking about 10-15s to analyze the full sequence CT volume. They asked the specialists to annotate the images after reading the patents' pathology report to increase the reliability of the training data. In the future, the authors will focus on dividing the lymph nodules into different medical situations and groups, as well as making use of context information in CT sequences.

**Nuclear segmentation** Phoulady, et al. [36] proposed a hierarchical multilevel thresholding method to segment nucleus. They firstly implemented color deconvolution according to Beer-Lambert Law and used hematoxylin channel grayscale image for nuclear segmentation. Then they binarized the image iteratively with the multilevel thresholding based on the Otsu method, in which case each subsequent binarization generated a subset of regions from the previous binarization. Visually, the segmentation result was inside the nuclei area, so they finally dilated the segmentation region with a 1-pixel radius disk. Their proposed method was tested on a variety of tissue and showed its generalization.

**Local and global panoramic view** Liu, et al. [37] proposed a method by using the local and global panoramic gastric view with gastroscopy to broaden the field of view (FOV) and provide doctors more sufficient information. A gastroscopy attached with a 6-degree of freedom (6-DOF) sensor was employed to obtain image sequence and position of camera. The fisheye effect of endoscope and the position relationship between scope and sensor was calibrated before clinical application. To realize the real-time local panoramic, the image was projected to a plane parallel to the 6-DOF position and the current displayed frame was stitched with the adjacent frame to broaden the FOV. The composited images were normalized in intensity channel in the HSI format to avoid generating apparent visual edge artifacts and blended with the multiresolution pyramidal algorithm followed by the cubic interpolation to solve the discontinuity.

As to obtain the global panoramic, the authors initially constrained the mosaicking with a dual-cubic to simplify and represent the complexed bean shape stomach. The dual-cubic model was built by optimizing the least square residual between the dual-cubic and the position of the selected landmark in stomach touched by the endoscopist during examination. The authors selected the frames that paralleled to the face of the dual-cubic to reduce distortion and selected proper overlay percentages to reduce computational cost and ensure accuracy of mosaicking process as well. They also implemented the bundle adjustment (BA) algorithm [41] to improve stitching accuracy. The local and global panoramic views were helpful in cancer diagnose and intraoperative navigation.

**3D reconstruction** Widya, et al. [38] reconstructed the 3D shape of a whole stomach from endoscopy videos and compared the performance with and without spraying indigo carmine color dye. Initially, camera calibration, frame extraction, and color channel separation were utilized to estimate the intrinsic camera parameters like focal length, projection center, and distortion parameters, which contributed to better reconstruction and corrected the distortion. Furthermore, the authors implemented the single channel input structure-from-motion (SfM) that extracted features with SIFT [25]. Subsequently, they matched the features to construct 3D points that were optimized by bundle adjustment [41]. The 3D points were meshed with the screened Poisson surface reconstruction [42] and applied with texture. Widya, et al. [38] discovered that the red channel of the video with IC blue dye achieved the best result.

**Cell tracking** The cell tracking in a video sequence seldomly yields good results, but is also useful in practice, such as investigating effect of different drugs by the movement trajectories of cells. Dorfer, et al. [39] proposed a method to recover the spatio-temporal cancer cell trajectories with the help of an approximate path annotated by a user watching the video sequence of cells, which did not contain temporal information. The automated cell detection method developed in the previous study [43] was used to detect the cells in a frame in the video. The closet points of the detected cells on the given approximate path were named as projections. Then a graph with a start point and an end point was formed, in which the nodes corresponding to the projections as well as the weighted edges were constructed based on the number of frames between the corresponding projections. The shortest path from start to end in the graph was determined by Dijkstra's algorithm. Finally, the path was re-parameterized as a one-dimensional function by the projections in the shortest path, where the temporal information was given by the order of frames. They could also determine a spatio-temporal path that went through the corresponding detected cells of the projections in the shortest path. The authors applied their method on the video of cultivated cancer cells on a matrix-coated glass-bottom culture dish and received satisfying results.

## V. DISSCUSION

### A. INSUFFICIENCY IN DATASETS IS THE BOTTLENECK FOR DEVELOPING POWERFUL MODELS

Datasets are the images with corresponding annotations. For one thing, the number of people who have gastric cancer or gastritis is limited and researchers can only have access to the images belong to their cooperative hospital. Hence, the number of accessible images is limited. For another, high quality annotation or labeling for images relies heavily on the experts' experience and is time consuming. Consequently, the difficulties of annotating images further exacerbate the insufficiency of datasets. The insufficiency of datasets has been a bottleneck for the development of powerful models to assist diagnosis and always results in the concern of model overfitting. There are some publicly available datasets, such as the dataset provided by the 2017 China Big Data and Artificial Intelligence Innovation and Entrepreneurship Competition with work [29], the pathological images dataset BOT with work [4], and the chromoendoscopy images dataset AIDA-E with works [12, 15].

### 1) PRODUCING MORE IMAGES WITH ANNOTATIONS

Data augmentation is easy but effective approach to enlarge the dataset by transformation like rotation, shifting, flipping, zooming, changing brightness, scale normalization and so on [2, 4, 5, 7-9, 11]. Data augmentation also helps to address overfitting caused by taking images under specific situation like in special angle or brightness. It is obvious that the annotation information is kept after transformation.

A GAN with a generation and a discriminator is a commonly adopted machine learning approach to enlarge the dataset via faked images with the same distribution as the real images. Basically, the generator tries to generate images look more and more real iteratively, while the discriminator tries to discriminate fake generated images from the real images iteratively. Under the adversarial situation, the generator can generate similar but new images, thus enlarging the dataset. To obtain the better annotation, Togo, et al. [1] adopted conditional GAN, and Kanayama, et al. [28] synthesized normal images and cancerous images at specified location.

### 2) DEVELOPING MORE POWERFUL MODELS ON THE LIMITED DATASETS

Transfer learning that utilized the model with parameters pretrained on a sufficient dataset on the insufficient medical dataset has the potential to perform well [5, 7, 9, 11, 35]. The approach is based on the observation that the shallow layers of deep learning models differ slightly, even if they are developed for different tasks. The powerful pretrained models are mostly developed from natural image datasets, like ImageNet. However, the efficiency of transfer learning from natural images still remain doubtful for the reason that natural images differ greatly from medical images [29].

Weakly supervised learning is an approach aimed at developing a powerful model with less annotation cost, which is easier to obtain. For segmentation, doctors need not to spend energy on annotating detailed boundary of cancer as ground truth [29]. MIL from CT sequence is almost annotation free, because it is confirmed that at least 1 abnormal image must exist in the CT sequence of sick people [2, 3]. Also, most of the medical images are labeled in image-level. It is useful to develop a weakly supervised approach to classify cropped patches with only the image-level annotation [10].

### B. FEATURE EXTRACTION AND SELECTION ARE CRUCIAL FOR BETTER MODELING

For image processing, features, a numeric vector, are needed to be extracted, to represent the input image in a high dimensional space for later handling the medical image processing tasks. There are two main kinds of feature extraction methods: One is the traditional hand-crafted methods, the other is the machine learning methods. There is a growing tendency of implementing machine learning techniques.

1) HAND-CRAFTED FEATURES

Hand-crafted features are always utilized with a specific filter or matrix, and rely on the researchers' intuition to calculate statistical representation. GLCM takes account of frequencies of gray level pixels appearing on a specific distance and angle [12, 14, 16]. Gabor filter is adopted to capture frequency information [3, 12, 14-16]. Principal component analysis (PCA) that carries out independent prime component with orthogonal transformation is frequently used for feature selection [17].

2) A GROWING ATTENTION IN MACHINE LEARNING METHODS

Convolutional neural network (CNN) is basically a dimension reduction method that reduces dimensionality in each layer after convoluting the former layers. CNN also has the ability to discover the relevance within a local area. The machine learning methods discussed in this paper are mostly developed based on CNN. With CNN, features can be extracted followed by a classifier, or we can directly input an image to CNN and output the classification result. Indeed, the convolutional layers in CNN extract features and the fully connected layers in CNN act as a classifier can be an explanation for that. However, the boundary of feature extraction then will be blurred. These years, a few of CNNs have well performance on processing natural images, like AlexNet [22], Inception-V3 [21], GoogLeNet [19], ResNet [24], DenseNet [23].

*C. SVM IS MOSTLY ONE OF THE BEST CLASSIFIERS*

As mentioned above, machine learning methods blur the boundary between features extraction and classification. But as an individual classifier, support vector machine (SVM) and random forest (RF) are always among the best ones. Moreover, SVM always reaches the highest performance [6, 12, 15-17].

*D. SEGMENTATION METHODS NEED MORE RESEARCHES*

Traditional region growing (RG), statistical region merging (SRM), Chan-Vese (C-V) model all carry out segmentation by applying statistical test or defining an intuitive function. Machine learning methods need to output pixel-level classification with the same size as the input image to get good segmentation results. To achieve this, fully convolutional network (FCN) that utilizes deconvolution operation and encoder-decoder based network that utilizes a decoder network to output pixel-level segmentation can be applied. The convolution operation in FCN and encoder in encoder-decoder based network can be regarded as feature extraction methods.

VI. CONCLUSION

Computer vision techniques are promising for the diagnosis of gastric cancer and gastritis for it is fast and labor-saving. Basically, lesion classification, cancer region detection and segmentation are the main tasks to find whether there are lesions or cancers, and where they are. Although the insufficiency of datasets has been an obstacle for developing powerful robust models for gastric image processing, machine learning methods usually outperform hand-crafted methods on feature extraction. GAN are implemented to enlarge the datasets and showed exciting performance. A number of methods, such as, CNN, Gabor filter, GLCM are developed to extract more representative features. Among the classifies, such as SVM, logistic regression, RF and naive Bayes, SVM is always the best classifier. Fully convolutional networks (FCN), region growing, Chan-Vese model, etc., are utilized to segment cancerous regions in images. Until now, more than a half of the research has focused on lesion classification, and more research is needed to focus on region detection and segmentation. Moreover, other approaches to broaden the doctor's view, to detect lymph nodules, to segment nuclei, and to track cells can also help doctors identifying gastric cancer more rapidly. According to the analysis result, as soon as more datasets made published or accessible across hospitals, computer vision assisted diagnosis will get into a new era that is indispensable for clinical doctors.

ACKNOWLEDGMENT

The author would like to thank Dr. Erwei Sun from The Third Affiliated Hospital of Southern Medical University for his critical review and advice.

TABLE I
PAPERS ON LESION CLASSIFICATION WITH ENDOSCOPY

| Paper | Categories | Methods | Database | Result |
|---|---|---|---|---|
| Zhao, et al. [18] (T) | 4 categories: gastric polyp, gastric ulcer, gastritis, and healthy | TirZ [18] and RF [a] | 574 gastrointestinal images (352×240) from 69 endoscopic videos | ACC=0.87 |
| Sakai, et al. [11] (T) | 2 categories: early gastric cancer and normal | GoogLeNet [19] | 926 images from 58 patients, with 228 images containing more than 1 lesion | SEN=0.800<br>SPE=0.948<br>ACC=0.876 |
| Liu, et al. [17] (T) | 2 categories: cancerous and normal | Joint diagonalization principal component discriminant analysis (JDPCDA), color coherence vector (CCV), and ODR-BSMOTE-SVM (OB-SVM) [20] | In total 1330 images from 291 patients, 768×576 for gastroscopy and 240×240 for wireless capsule endoscopy (WCE) | 1) For gastroscopy:<br>SEN=0.9077<br>SPE=0.9074<br>ACC=0.9075<br>AUC=0.9532<br>2) For WCE:<br>SEN=0.9385<br>SPE=0.9450<br>ACC=0.9434<br>AUC=0.9776 |
| Ali, et al. [12] (E) | 2 categories: normal and abnormal | Gabor-based GLCM [b] (G2LCM), SVM [d] | 176 chromoendoscopy frames (publicly available from AIDA-E) | SEN=0.91<br>SPE=0.82<br>ACC=0.88<br>AUC=0.91 |
| Boschetto and Grisan [13] (E) | 2 categories: normal and abnormal (metaplasia and dysplasia lesions) | Simple linear iterative clustering (SLIC), GLCM [b], LBP [c], and RF [a], majority vote | 176 images selected from examinations of 28 patients | ACC=0.9205 |
| Ali, et al. [15] (E) | 2 categories: normal and abnormal (metaplasia, dysplasia) | Gabor filter, genetic algorithm (GA), SVM [d] | 176 chromoendoscopy frames (publicly available from AIDA-E) | SEN=0.927<br>SPE=0.887<br>ACC=0.915<br>AUC=0.94 |
| Liu, et al. [5] (E) | 2 categories: normal and early gastric cancer | Inception-V3 | 719 early gastric cancer and 1612 normal images of 4 resolution, 768×576, 720×480, 1920×1080, 1980×720 | ACC=0.985<br>SEN=0.981<br>SPE=0.989 |

[a] random forest (RF), [b] gray level co-occurrence matrix (GLCM), [c] local binary pattern (LBP), [d] support vector machine (SVM)

TABLE II
PAPERS ON LESION CLASSIFICATION WITH PATHOLOGICAL IMAGES

| Paper | Categories | Methods | Database | Result |
|---|---|---|---|---|
| Li, et al. [4] | 2 categories: normal and cancerous | GrastricNet based on CNN [a] | Publicly available BOT dataset containing 560 cancerous and 140 normal slices (2048×2048) | For patch-based:<br>ACC=0.9793<br>For slice-based:<br>ACC=1 |
| Li, et al. [10] | 2 categories: normal and abnormal | VGG-16 [27], ANN [b], multilayer hidden conditional random fields (MHCRF) | 100 normal and 100 abnormal images (2048×2048) | ACC=0.93<br>SEN=0.92<br>PRE=0.9388<br>SPE=0.94<br>FSC=0.9293 |
| Sharma, et al. [7] | 1) 3 categories for cancer: HER2+ tumor, HER2- tumor, non-tumor;<br>2) 2 categories for necrosis: necrosis and non-necrosis | CNN [a] | Whole slide images of surgical sections selected from 454 cases of gastric adenocarcinoma | 1) For cancer:<br>ACC=0.6990<br>2) For necrosis:<br>ACC=0.8144 |
| Kong, et al. [8] | 2 categories: invasive gastric cancer or normal | Depthwise separable convolution and FCN [c] | Whole slide images containing 146 normal and 126 tumors (average testing size 107595×161490) | FROC=0.811 [d] |

standard imaging endoscopy (T), image-enhanced endoscopy (E)
[a] convolutional neural network (CNN), [b] artificial neural network (ANN), [c] fully convolutional networks (FCN), [d] Free-response ROC (FROC)

TABLE III

TABLE III
PAPERS ON LESION CLASSIFICATION WITH X-RAY

| Paper | Categories | Methods | Database | Result |
|---|---|---|---|---|
| Togo, et al. [1] | 2 categories: gastritis and non-gastritis | Loss function-based conditional progressive growing GAN (LC-PGGAN) | 100 gastritis and 100 non-gastritis images are collected to train for image generation, and 615 images are collected as test data | - |
| Ishihara, et al. [6] | 2 categories: having and not having gastric cancer risk | CNN [a], SVM [b] | 1024×1024 X-ray images of 8-bit gray scale from 2100 patients where 971 are considered to have gastric risk | SEN=0.895<br>SPE=0.935<br>HAR=0.914 [c] |
| Togo, et al. [16] | 2 categories: chronic gastritis and non-gastritis | SVM [b] | 1957 gastric X-ray images (1024×1024) | ACC=0.775<br>SEN=0.814<br>SPE=0.741<br>YDI=0.555 [e] |
| Ishihara, et al. [14] | 2 categories: helicobacter pylori infected and uninfected | mRMR [d], SVM [b], multiple kernel learning (MKL), expectation maximization (EM) algorithm | 2100 samples including 1056 males and 1044 females | SEN=0.895<br>SPE=0.896<br>HAR=0.895 [c]<br>AUC=0.943 |

[a] convolutional neural network (CNN), [b] support vector machine (SVM), [c] harmonic mean of sensitivity and specificity (HAR), [d] minimum-redundancy maximum-relevance algorithm (mRMR), [e] Youden index (YDI)

TABLE IV
PAPERS ON LESION CLASSIFICATION WITH COMPUTED TOMOGRAPHY (CT)

| Paper | Categories | Methods | Database | Result |
|---|---|---|---|---|
| Li, et al. [3] | 2 categories: normal and cancerous | MIL [a], Circular Gabor filter (CGF), Citation-KNN [c] | 26 patients with gastric cancer invaded | SEN=0.8750<br>FPR=0.4000 [d]<br>ACC=0.7692 |
| Fang, et al. [2] | 1) 2 different degrees: poorly and well/moderately;<br>2) Laurent classification: intestinal, diffuse, and mixed | deep generalized MIL [a] (GMIL) based on CNN [b] | CT volumes including 433 gastric cancer patients | 1) For 2 different degrees: ACC=0.815<br>2) For Laurent classification: ACC=0.731 |
| Fang, et al. [9] | Discriminating serosa invasion (T4 stage) in advanced gastric cancer | Multi-task learning, AlexNet [22] | 357 advanced gastric cancer patients with 153 T3 and 204 T4 | AUC=0.842<br>ACC=0.793<br>SEN=0.820<br>SPE=0.756 |

[a] multi-instance learning (MIL), [b] convolutional neural network (CNN), [c] k-nearest neighbor (KNN), [d] false positive rate (FPR)

TABLE V
PAPERS ON CANCER REGION DETECTION AND SEGMENTATION

| Paper | Image type | Methods | Database | Result |
|---|---|---|---|---|
| Kanayama, et al. [28] (D) | Endoscopic images | Generative adversarial networks (GAN) | 129692 normal images and 1315 lesion images which includes 1309 tumors and 6 ulcers | - |
| Ural, et al. [34] (D) | CT images | Power Spectrum Density (PSD) analysis | CT scans of 30 patients from above and opposite sides | Success rate=0.83 [a] |
| Sun, et al. [30] (S) | Pathological images | ResNetV2-101 based encoder-decoder | 500 images(2048×2048) containing typical cancerous region | ACC=0.9160<br>AUC=0.8265 |
| Liang, et al. [29] (S) | Bionomical images | Fully convolutional networks (FCN) | 1400 weakly annotated images (2018×2048), 400 precisely labeled validation images and 100 test images | ACC=0.9109<br>IOU=0.883 [b] |
| Yasar, et al. [31] (S) | Endoscopic images | Region growing (RG) and statistical region merging (SRM) | patients who visited the gastroenterology department during a year since June 2016 | ACC=0.9633<br>SEN=0.8501<br>SPE=0.9772<br>PRE=0.9268<br>FSC=0.8853 |
| Wang, et al. [32] (S) | Polarimetry imaging images | Mueller matrix | 84 samples from 59 patients, including 20 cancer, 15 dysplasia, 22 intestinal metaplasia, and 27 normal glands samples. | ACC=0.59 |
| Liu, et al. [33] (S) | M-NBI [c] images | Chan-Vese model and hue texture | 740 M-NBI [c] images (768×576) from 49 patients | FSC=0.61<br>FPR=0.16 [d] |

Cancer region detection (D), cancer region segmentation (S)
[a] the rate of successful detection claimed by the authors, [b] intersection over union (IOU), [c] magnifying endoscopy with narrow-band imaging (M-NBI), [d] false positive rate (FPR)

TABLE VI
PAPERS ON LESS RESEARCHED BUT IMPORTANT QUESTIONS

| Paper | Image type | Achievement | Methods | Database | Result |
|---|---|---|---|---|---|
| Zhengdong, et al. [35] | CT images | Metastatic lymph nodules detection | Faster R-CNN [40] | 2400 images of 512×557 | PRE=0.9231<br>SEN=0.6471<br>SPE=0.9752<br>ACC=0.8324<br>AUC=0.9248 |
| Phoulady, et al. [36] | Histological images | Nucleus segmentation | Color deconvolution, image reconstruction, multilevel thresholding | 36 histology images from a wide variety of tissues including 7931 labeled cells | PRE=0.929<br>SEN=0.886<br>FSC=0.907 |
| Liu, et al. [37] | Endoscopic images | Provide postoperative global and real-time local panoramic view from gastroscopy | Frame stitching, multiresolution pyramidal algorithm, bundle adjustment [41] | Simulated stomach phantom and in-vivo data | 7.12 frames/s in standard computer environment. Mosaicking error is 0.43mm and 3.71mm for local and global panoramic view respectively. |
| Widya, et al. [38] | Endoscopic images | 3D reconstruction of whole stomach | Structure-from-motion (SfM), bundle adjustment [41], screened Poisson surface reconstruction [42] | AVI format 30 frames/s full HD resolution videos containing 2 image sequence with and without spraying dye for each video | Over 96% images are reconstructed |
| Dorfer, et al. [39] | Phase contrast video-microscopy images | Tracking cancer cells for spatio-temporal trajectories | Detection-graphs and shortest path searching | 7 video sequences in which images are of 512×512 | Recover 92% of the ground truth trajectory points |